\pdfoutput=1
\documentclass[a4paper,10pt,amsmath,amssymb,notitlepage,twoside,superscriptaddress,nofootinbib,eqsecnum,aps,prd]{revtex4-1}
\usepackage[latin1]{inputenc}
\usepackage[T1]{fontenc}
\usepackage[usenames]{color}
\usepackage[final]{showkeys} 
\usepackage{enumerate}
\usepackage[pdftex]{hyperref}
\hypersetup{colorlinks=true,linkcolor=black,citecolor=black,filecolor=black,urlcolor=black,
            pdfauthor={M. Cadoni, E. Franzin and M. Serra},
            pdftitle={Brane solutions sourced by a scalar with vanishing potential and classification of scalar branes}}
\usepackage[capitalise]{cleveref} 
\crefformat{plural}{#2Eqs.~(#1)#3} 
\crefname{section}{Sect.}{Sects.}

\makeatletter\g@addto@macro\bfseries{\boldmath}\makeatother%

\newcommand*{\refcite}{Ref.~\cite}

\renewcommand{\geq}{\geqslant}

\renewcommand{\ge}{\geqslant}
\renewcommand{\le}{\leqslant}

\def\0{\nonumber}
\def\AdS{\text{AdS}}
\def\lb{\label}

\def\eg{\emph{e.g.}}
\def\ie{\emph{i.e.}}
\def\l{\lambda}

\def\p{\partial}
\DeclareMathOperator\const{const}

\def\be#1\ee{\begin{align}#1\end{align}}
\def\bsube[#1]#2\esube{\begin{subequations}\label[plural]{#1}\begin{align}#2\end{align}\end{subequations}}

\begin{document}

\title{\texorpdfstring{Brane solutions sourced by a scalar with vanishing potential\\and classification of scalar branes}{Brane solutions sourced by a scalar with vanishing potential and classification of scalar branes}}
\author{Mariano Cadoni}\email{mariano.cadoni@ca.infn.it}
\affiliation{Dipartimento di Fisica, Universit\`a di Cagliari,\\Cittadella Universitaria, 09042 Monserrato, Italy}
\affiliation{INFN, Sezione di Cagliari}
\author{Edgardo Franzin}\email{edgardo.franzin@ca.infn.it}
\affiliation{Dipartimento di Fisica, Universit\`a di Cagliari,\\Cittadella Universitaria, 09042 Monserrato, Italy}
\affiliation{INFN, Sezione di Cagliari}
\affiliation{CENTRA, Departamento de F{\'\i}sica, Instituto Superior T\'ecnico, Universidade de Lisboa,\\Avenida Rovisco Pais 1, 1049 Lisboa, Portugal}
\author{Matteo Serra}\email{serra@mat.uniroma1.it}
\affiliation{Dipartimento di Matematica, Sapienza Universit\`a di Roma\\Piazzale Aldo Moro 2, 00185 Roma, Italy}
\date{\today}

\begin{abstract}
We derive exact brane solutions  of minimally coupled Einstein-Maxwell-scalar gravity 
in $d+2$ dimensions with a vanishing scalar potential. The solutions  are conformal to the Lifshitz
spacetime and have several properties, which make them interesting  for holographic applications.
In particular, the dual QFT is characterized by hyperscaling  violation.
We show that our solutions, together with the AdS brane and the domain wall sourced by an exponential
potential, give the complete  list of  scalar  branes  sourced by a generic potential having simple
(scale-covariant)  scaling symmetries. This allows us to give a complete classification of both simple
and interpolating brane solution of minimally coupled Einstein-Maxwell-scalar gravity,
which may be very useful for holographic applications.
\end{abstract}

\maketitle

\section{Introduction}

In its  most general setting, Einstein-Maxwell-scalar gravity (EMSG) is characterized by three coupling functions:
the self-interaction scalar potential $V(\phi)$, the gauge coupling function $Z(\phi)$, and the coupling function $S(\phi)$
responsible for the mass of the Maxwell field. In recent times, a lot of effort has been devoted to the derivation and investigation
of brane solutions in the context of EMSG in $d+2$ dimensions with various forms of the coupling functions~\cite{Charmousis:2009xr,%
Gubser:2009qt,Goldstein:2009cv,Cadoni:2009xm,Bertoldi:2011zr,Cadoni:2011nq,Gouteraux:2011qh,Cadoni:2011yj,Cadoni:2012uf,%
Cadoni:2012ea,Gouteraux:2012yr,Cadoni:2013hna}.

The main reason behind  this interest is  the holographic application  of this kind of solutions. 
Using the rules of the AdS/CFT correspondence~\cite{Maldacena:1997re,Horowitz:2006ct,Hartnoll:2009sz},
one can investigate the QFT dual to the brane and in particular derive its transport properties.
This strategy  has allowed to uncover  a very rich  phenomenology ranging from applications to condensed matter
systems---holographic superconductors~\cite{Hartnoll:2008vx,Horowitz:2008bn,Herzog:2009xv,Horowitz:2010gk},
phase transitions~\cite{Cadoni:2009xm,Cadoni:2011kv}, quantum criticality~\cite{Gubser:2008wz,Gouteraux:2011ce,Gouteraux:2012yr},
hyperscaling violation in critical systems~\cite{Cadoni:2012ea,Cadoni:2012uf,Dong:2012se,Narayan:2012hk,Perlmutter:2012he,Ammon:2012je,%
Bhattacharya:2012zu,Alishahiha:2012cm,Dey:2012fi,Sadeghi:2012vv,Alishahiha:2012qu,Kim:2012pd,Edalati:2012tc,Gath:2012pg,Cremonini:2012ir,%
Hassaine:2015ifa,Bravo-Gaete:2015wua,Kofinas:2015zaa}, fluidodynamics~\cite{Salvio:2012at,Roychowdhury:2015cta},
entanglement entropy~\cite{Ogawa:2011bz,Narayan:2012ks,Narayan:2013qga,Cremonini:2013ipa}---to
cosmology~\cite{Skenderis:2006jq,Shaghoulian:2013qia,Cadoni:2013gza,Mignemi:2014kea,Cadoni:2015iua}.
However, the essential qualitative features of the dual QFT are not pertinent
to the presence of non-trivial coupling functions $Z$ and $S$ but  are basically determined by the presence of a non-constant
scalar field. In the dual QFT the non-trivial profile of the scalar field plays the crucial role of an order parameter
triggering symmetry breaking and/or phase transitions.  For this reason, in this paper we will restrict our considerations
to \emph{minimally coupled} Einstein-Maxwell-scalar gravity (MCEMSG), which is characterized by a single coupling function, the potential
$V(\phi)$, whereas the other two are trivial, \ie\ $Z=1$ and $S=0$.

The branes  of MCEMSG have been investigated in several papers and both analytical  and numerical solutions have been
derived~\cite{Cadoni:2009xm,Charmousis:2009xr,Goldstein:2009cv,Gubser:2009qt,Cadoni:2011yj,Cadoni:2011nq,Cadoni:2012ea,%
Gouteraux:2011qh,Bertoldi:2011zr,Cadoni:2012ea,Gouteraux:2012yr,Cadoni:2012uf,Cadoni:2013hna}. There are three different types of solutions.
\emph{Black branes} (BB) are solutions with a singularity shielded by an event horizon. In the dual QFT these solutions
correspond to a field theory at finite temperature $T$ and are very important in the description of phase
transitions~\cite{Hartnoll:2008vx,Horowitz:2008bn,Herzog:2009xv,Horowitz:2010gk,Cadoni:2009xm,Cadoni:2011kv,Cadoni:2012uf}.
\emph{Scale-covariant branes} (SCB) are a sort of elementary solutions transforming  covariantly under scale transformations.
They represent a generalization of the usual Minkowski or AdS vacua to spacetimes with non-standard asymptotics~\cite{Cadoni:2011nq,Dong:2012se,Cadoni:2013yxa}.
They are sourced by a scalar field with $\log r$ behavior, have no horizon and in the dual QFT  typically  correspond
to a $T=0$ ground state exhibiting hyperscaling violation~\cite{Cadoni:2012ea,Cadoni:2012uf,Cremonini:2012ir,Bhattacharya:2012zu,%
Dong:2012se,Narayan:2012hk,Gath:2012pg,Ammon:2012je,Alishahiha:2012cm,Alishahiha:2012qu,Perlmutter:2012he,Dey:2012fi,Sadeghi:2012vv}.
The AdS brane, which is characterized by full conformal invariance and is sourced by a constant scalar, appears as limiting case
of this class of solutions. Scale-covariant solutions  in general  have a curvature singularity at $r=0$ and need therefore an
infrared (IR) regularization~\cite{Bhattacharya:2012zu}. \emph{Interpolating branes} are solutions which interpolate between two
elementary branes  at $r\sim0$ (the IR of the dual QFT) and $r=\infty$ (the ultraviolet (UV) of the dual QFT). They are $T=0$ solutions
describing the flow of the dual QFT from a IR to an UV regime, in which the solution behaves as an elementary, scale-covariant, solution. 

The  presence of a given type of brane solution in the  spectrum of  MCEMSG theories depends on the specific  form of the potential 
$V(\phi)$. Elementary, scale-covariant branes require an exponential potential (a constant $V$ is required for the AdS brane),
whereas black or interpolating branes typically require a more complicate potential with different behaviour in the $r=0$
and $r=\infty$ region. Notice that the known no-hair theorems~\cite{Israel:1967wq,Bekenstein:1995un,Hertog:2006rr} only apply to
asymptotically  flat or asymptotically AdS solutions. Thus, solutions with non-standard asymptotics do not necessarily satisfy
no-hair theorems.  Although several solutions of  MCEMSG theories are known, presently it is not clear if the scale-covariant geometries
found until now exhaust all the possible solutions of this kind of theory.
Clearly, this lack of knowledge prevents a complete classification of the possible interpolating geometries.

The purpose of this paper is twofolds. First, we derive the exact solutions of MCEMSG in the case of a vanishing potential,
which are the brane counterpart  of the Janis-Newmann-Winicour-Wyman (JNWW) solutions~\cite{Janis:1968zz,Wyman:1981bd,Cadoni:2015gfa,Cadoni:2015qxa}.
We show that these solutions belong to the class of scale-covariant  solutions generating hyperscaling violation in the dual QFT\@.
Second, we demonstrate that these solutions complete the list of the possible scale-covariant solutions of the theory.
This will allow us to give  an  exhaustive  classification of the interpolating brane solutions of MCEMSG\@. 

The structure of this paper is the following. In \cref{sect:s2} we briefly review MCEMSG and the parametrization of the field equations
introduced in \refcite{Cadoni:2011nq}. In \cref{sect:hyper} we classify branes according to their hyperscaling violation parameter and dynamic scaling exponent.
We then derive and discuss the brane solutions of the theory with a vanishing potential in \cref{sect:s3}.
In \cref{sect:s4}  we show that this solution completes the list of elementary, scale-covariant branes  of the theory
and we give a complete classification of both elementary and interpolating solutions.  In \cref{sect:s4_1} we construct
an explicit example of interpolating solution, not yet discussed in the literature. Finally, we conclude in \cref{sect:concl}.
In \cref{app:chargedJNWW} we present a result, which is somehow beyond the scope of this paper, but can be obtained  using our
parametrization of the field equations, namely the charged, spherically symmetric solutions of MCEMSG\@.
These solutions are the charged generalization of the JNWW solutions.

\section{Einstein-Maxwell-scalar gravity\lb{sect:s2}}

We consider MCEMSG in $d+2$ spacetime dimensions (with $d\geq 2$):
\be\label{action}
A=\int{}d^{d+2}x\,\sqrt{-g}\left(\mathcal{R} -2 (\p\phi)^2 -F^2 -V (\phi)\right),
\ee%
where $\mathcal{R}$ is the scalar curvature of the spacetime. The ensuing field equations are
\bsube[max_scal_eq]
\nabla_\mu{}F^{\mu\nu}&= 0,\\
\nabla^2\phi&= \frac{1}{4}\frac{dV(\phi)}{d\phi},\\
\mathcal{R}_{\mu\nu} -\tfrac{1}{2}g_{\mu\nu}\mathcal{R}&= 2\left(F_{\mu\rho}F_\nu^\rho-\tfrac{1}{4}g_{\mu\nu} F^{\rho\sigma}F_{\rho\sigma}\right)
+2\left(\p_\mu\phi\,\p_\nu\phi -\tfrac{1}{2}g_{\mu\nu} \p^\rho\phi\,\p_\rho\phi\right) -\tfrac{1}{2}g_{\mu\nu} V(\phi).
\esube%

We are interested in static, radially  symmetric solutions of the field equations,
for this we parametrize the  spacetime metric in a Schwarzschild gauge: 
\be\label{pmetric}
ds^2 = -U (r)\,dt^2 +U^{-1} (r)\,dr^2 +R^2 (r)\,d\Omega_{(\varepsilon,d)}^2,
\ee%
where $\varepsilon=0,1,-1$ denotes, respectively, the $d$-dimensional
planar, spherical, or hyperbolic transverse space with metric $d\Omega_{(\varepsilon,d)}^2$. 
Moreover, we will consider only
purely electric solutions; magnetic solutions can be easily generated from the electric ones using the electro-magnetic duality.
Let $Q$ be the electric charge. Then, the electric field satisfying~\eqref{max_scal_eq} reads 
\be%
F_{tr}=\frac{Q}{R^d}.
\ee%
Under these assumptions, the field equations take the following form 
\bsube[fed]
\frac{R''}{R} &=-\frac{2}{d}(\phi')^2,\\
(UR^d\phi')' &=\frac{1}{4}R^d\frac{dV}{d\phi},\\
(UR^d)'' &=\varepsilon{} d(d-1)R^{d-2}+2\frac{d-2}{d}\frac{Q^2}{R^d}-\frac{d+2}{d}R^{d}V,\\
(UR^{d-1}R')' &=\varepsilon{} (d-1)R^{d-2}-\frac{2}{d}\frac{Q^2}{R^d}-\frac{1}{d}R^{d}V.
\esube%

In general, the solutions of these field equations depend on the class of potentials $V(\phi)$ considered.
Usually one must impose precise boundary conditions on the $r=\infty$ asymptotic behavior of the solution, which translate
into boundary conditions  for the potential $V$. For example, if we require an asymptotically flat spacetime and assume
without loss of generality that $\phi(r\to\infty)=0$, it follows that $V(0)=0$, while for asymptotically AdS
spacetimes we have $V(0)=-\Lambda^2$.
Typically one also requires the existence of the Schwarzschild black hole (black brane) solution sourced by a constant scalar
field $\phi=0$, implying $V'(0)=0$, while the existence of black hole solutions sourced by a non-trivial scalar field in general
is strongly constrained by well-known no-hair theorems.

In general, finding exact solutions of the field equations~\eqref{fed} is a very difficult task even when the explicit form of
the potential $V$ is given. An efficient solving method has been proposed in \refcite{Cadoni:2011nq}.
Such method is particularly useful in the holographic context and it has been successfully used in several cases to generate
exact, asymptotically flat or AdS, solutions of Einstein-(Maxwell)-scalar gravity, where the potential is not an input
but an output of the theory~\cite{Cadoni:2011nq,Cadoni:2015gfa,Cadoni:2015qxa}.

Following \refcite{Cadoni:2011nq}, introducing the variables $R=e^{\int{}Y}$ and $u=UR^d$, the field equations~\eqref{fed} become 
\bsube[zzll]
Y'+Y^2&=-\frac{2}{d} (\phi')^2,\label{z0}\\
(u\phi')' &=\frac{1}{4}e^{d\int{}Y}\frac{dV}{d\phi},\label{z1}\\
u''-(d+2)(uY)' &=-2\varepsilon{} (d-1)e^{(d-2)\int{}Y}+4Q^2e^{-d\int{}Y},\label{z2}\\
\quad u'' &=\varepsilon{} d(d-1)e^{(d-2)\int{}Y}+2\frac{d-2}{d}Q^2e^{-d\int{}Y}-\frac{d+2}{d}e^{d\int{}Y}V.\label{z3}
\esube%
\cref{z2,z3} are second-order  linear differential equations in $u$, whereas~\eqref{z0} is a first-order nonlinear equation
for $Y$, known as the Riccati equation. In general, starting from a given profile $\phi(r)$ for the scalar field, one can look for solutions
of the Riccati equation. Once the solution for $Y$ has been found, one can integrate \cref{z2}, which is linear in $u$, to obtain
\be\label{sol1}
u=R^{d+2}\left[\int\left(4Q^2\int\frac{1}{R^d} -2\varepsilon{} (d-1)\int R^{d-2} -C_1\right)\frac{1}{R^{d+2}} +C_2\right],  
\ee%
where $C_{1,2}$ are integration constants. The last step is the  determination of  the potential using \cref{z3}:
\be\label{potential}
V=\frac{d^2(d-1)}{d+2}\frac{\varepsilon}{R^2}+2\frac{d-2}{d+2}\frac{Q^2}{R^{2d}}-\frac{d}{d+2}\frac{u''}{R^d}.
\ee%

\section{Scaling symmetries and hyperscaling violation\lb{sect:hyper}}

The $d+2$ dimensional metric for a brane ($\varepsilon=0$), whose dual QFTs are characterized by hyperscaling violation  is usually written as~\cite{Dong:2012se}
\be\lb{hv}
ds^2=  r^{-2(d-\theta)/d}\left(-r^{-2(z-1)}\,dt^2 + dr^2 + dx_i\,dx^i\right),
\ee%
where $dx_i\,dx^i=d\Omega_{(0,d)}^2$,  $\theta$ is the hyperscaling violation parameter and $z$ is the  dynamic scaling exponent characterizing the anisotropic scaling 
of the time and space coordinates, which breaks Lorentz invariance in the dual QFT\@. The scaling symmetries of the metric~\eqref{hv} are
\be\lb{ss}
t\to\l^{z}t,\quad r\to\l{}r,\quad 
x_i\to\l{}x_i,\quad ds\to\l^{\theta/d}ds.
\ee%
It follows that a nonzero value of $\theta$ makes the metric~\eqref{hv} not  scale-invariant,
but only  \emph{scale-covariant}, in the sense that the metric transforms with a definite weight under a scale transformation.
The scaling transformation determines the following scaling behavior for the free energy of the dual QFT, given in terms of $\theta$ and $z$:
\be\lb{fe}
F\sim{}T^{\frac{d+z-\theta}{z}}.
\ee%

It is useful to distinguish between the two effects of hyperscaling violation ($\theta\neq0$) and 
anisotropic scaling ($z\neq 1$) introducing  four different subclasses:
\begin{enumerate}[(1)]
\item\lb{item1} $\theta=0, z=1$ describes \emph{AdS  branes}. The metric~\eqref{hv} gives the  AdS spacetime in $d+2$ dimensions,
the scaling~\eqref{ss} is isotropic and the dual QFT is a CFT\@.
\item\lb{item2} $\theta=0, z\neq 1$ describes \emph{Lifshitz branes}. Because $\theta=0$, the metric is not only covariant but also
invariant under the scale transformation~\eqref{ss}. On the other hand, because $z\neq 1$ the scaling is not isotropic in the
$t$ and $x_i$ coordinates and the dual $d+1$-dimensional QFT is not invariant under the $d+1$-dimensional Lorentz group.
\item\lb{item3} $\theta\neq 0, z=1$ describes \emph{Domain Walls}. The scaling~\eqref{ss} is isotropic but being $\theta\neq0$
full scale invariance is broken and only scale covariance survives. The metric is conformal to AdS spacetime. The dual QFT is Lorentz
invariant and one can still formulate a DW/QFT correspondence~\cite{Boonstra:1998mp,Kanitscheider:2009as}. Notice that Minkowski
spacetime in $d+2$ dimensions is a particular case of this class of solutions, it is obtained for $\theta=d$.
\item\lb{item4} $\theta\neq 0, z\neq1$ describes \emph{Conformal-Lifshitz branes}. Now the scaling~\eqref{ss} is anisotropic
and hyperscaling is violated. In this case the metric~\eqref{hv} is conformal to the Lifshitz spacetime~\cite{Cadoni:2011nq}.
\end{enumerate}

\section{\texorpdfstring{Brane solutions sourced by a scalar field with $V=0$}{Brane solutions sourced by a scalar field with V=0}\lb{sect:s3}}

The form~\eqref{zzll} of the field equations is also very useful for deriving exact solutions  in the case of a identically
vanishing potential, $V(\phi)=0$. This is an important particular case in the context of Einstein-scalar models. In particular, electrically
neutral ($Q=0$), spherically symmetric ($\varepsilon=1$) black hole solutions of the \cref{max_scal_eq} are known as the
JNWW solutions~\cite{Janis:1968zz,Wyman:1981bd}. They are asymptotically flat solutions, have naked singularities (consistently
with the known no-hair  theorems) and present several interesting features~\cite{Cadoni:2015gfa,Cadoni:2015qxa}.

In this paper we will focus our attention on solutions in $d+2$ spacetime dimensions 
for  which $\varepsilon=0$ (branes), \ie\ the transverse 
$d$-dimensional sections have planar topology. In this case the field equations~\cref{zzll} reduce to
\bsube[z3LL]
&Y'+Y^2=-\frac{2}{d} (\phi')^2,\label{z11}\\
& (u\phi')'=0,\label{z12}\\
&u''- (d+2) (uY)'=4Q^2 e^{-d\int{}Y},\label{z13}\\
&u'' =2\frac{d-2}{d} Q^2 e^{-d\int{}Y}\label{z14}.
\esube%
Let us discuss separately the uncharged ($Q=0$) and charged case.

\subsection{Uncharged brane solutions\lb{sect:s31}}

We consider first the electrically neutral case $Q=0$.
One can easily realize that flat branes with $R=r$ and $U=1$ (corresponding to  $u=r^d,\, Y=1/r$), are not solution
of the field equations~\eqref{z3LL}. However, the system~\eqref{z3LL} can be integrated exactly and we can
find the most general form of the solution. In fact, we can directly solve the trivial equation~\eqref{z14}, giving $u$
as a linear function of $r$. Then we solve \cref{z13} for $Y$, and finally determine $\phi$ using \cref{z12}. 
The Riccati equation then gives just a constraint for the integration constants. We find:
\be\lb{sol31}
U=\left(\frac{r}{r_0}\right)^{1-dw},\quad
R^2=\left(\frac{r}{r_0}\right)^{2w},\quad
\phi=-\gamma\log\left(\frac{r}{r_0}\right)+\phi_0,\quad 
w-w^2=\frac{2}{d}\,\gamma^2,
\ee%
where $r_0,\gamma,w,\phi_0$ are integration constants. The constraint implies the condition $0\le{}w\le1$. The solution~\eqref{sol31} is invariant under the
transformation $w\to1-w$, which maps solutions with $w\in[0,1/2]$ into solutions with $w\in [1/2,1]$.
Neglecting the constant $\phi_0$,  a trivial translation mode of the scalar,  these solutions give a two-parameter
($r_0, w$) family of  brane solutions. In particular, $r_0$ represents a length scale, while $w$ is a dimensionless parameter.
The solution~\eqref{sol31} has interesting  scaling symmetries, which  we discuss in detail in \cref{subsect:hyper}.
Moreover, it  can be considered as the brane counterpart of the JNWW spherical solutions.

For $w\neq 0,1$ the solutions have a naked singularity at $r=0$, in fact the scalar curvature is 
\be\lb{sc}
\mathcal{R}=\frac{2\gamma^2}{r_0^2}\left(\frac{r}{r_0}\right)^{-1-dw}.
\ee%
Well-known no-hair theorems~\cite{Israel:1967wq,Bekenstein:1995un,Hertog:2006rr}  forbid asymptotically flat BB
solutions when $V=0$. In principle these theorems  do not apply to solutions with non-standard asymptotics  such as~\eqref{sol31}.
Nevertheless, one can easily check that \cref{z3LL} do not allow for solutions with event horizons. Thus,
the brane~\eqref{sol31} is a $T=0$ solution, which does not allow for finite temperature excitations.

In the two limiting cases $w=0,1$ we  have a constant scalar and the corresponding brane solutions  become respectively 
\be\lb{kk1}
ds^2= -\frac{r}{r_0}dt^2+ \frac{r_0}{r}dr^2+ dx_i\,dx^i,\quad
ds^2= -\left(\frac{r}{r_0}\right)^{1-d}dt^2+\left(\frac{r}{r_0}\right)^{d-1}dr^2+ \left(\frac{r}{r_0}\right)^2dx_i\,dx^i.
\ee%

The  $w=0$ brane is just $d+2$-dimensional Minkowski space in a particular coordinate system. In fact, the $(r,t)$
sections of the metric can be brought in the Rindler  form by  a simple redefinition of the radial coordinate $r$.
The $w=1$ brane describes instead a Ricci-flat manifold, which can be considered as the brane counterpart of the Schwarzschild black hole.

\subsubsection{Energy of the brane}

Let us now calculate the total energy $M$  of the solution, using the Euclidean action formalism of \refcite{Martinez:2004nb}.
The variation of the boundary terms of the action consists of both a gravitational and a scalar contribution:
\be\label{mass}
\delta{}M = 8\pi\left[-\frac{d}{2} R^{d-1}R'\delta{}U +U' R^{d/2}\delta\left(R^{d/2}\right)
-d\,U R^{d/2}\delta{}\left(R^{\frac{d-2}{2}}R'\right)\right]^\infty -16\pi\left[R^{d}U\phi'\delta\phi\right]^\infty.
\ee%
Evaluating this equation for the solution~\eqref{sol31} one finds
\be\label{mass0}
M=\frac{4\pi}{r_0}dw[(2-d)w-2].
\ee%

Taking into account the constraints on $w$ and $d$ (namely $0\le{}w\le1$ and $d\geq2$), the energy vanishes only
when $w=0$, \ie\  for Minkowski space, while for $0<w\le1$ the sign of the energy is ruled by the sign of $r_0$
(negative for $r_0>0$ and positive for $r_0<0$). Taking $r>0$,  solution~\eqref{sol31} exists only for $r_0>0$,
thus the brane has always negative mass. This behaviour is quite different from the JNWW solutions, where the sign
of the energy depends both on the value of the dimensionless parameter of the solutions and on $r_0$~\cite{Cadoni:2015gfa,Cadoni:2015qxa}.

\subsubsection{Scaling symmetries and hyperscaling violation\lb{subsect:hyper}}

The brane metric~\eqref{sol31} has remarkable scaling symmetries and can be put in the form~\eqref{hv} by the transformation of coordinates
\be\lb{ct}
\frac{r}{r_0}\to\left(\frac{r}{\tilde{r}_0}\right)^{\frac{2}{1+w (d-2)}},\quad\tilde{r}_0=\frac{2r_0}{1+w(d-2)}.
\ee%

The hyperscaling violation parameter and the dynamical exponent are given by
\be\lb[plural]{para}
z = \frac{2d w}{1+w (d-2)},\quad\theta=\frac{d(1+dw)}{1+w (d-2)}.
\ee%
We observe that $z$ and $\theta$ are not independent but satisfy the relation
\be\lb{rel}
\theta=z+d.
\ee%

From \cref{para} we see that the brane solution~\eqref{sol31} never belongs to the subclasses (\ref{item1}) or (\ref{item2}) of \cref{sect:hyper},
\ie\ it can neither describe an AdS spacetime nor a Lifshitz spacetime. 
It can be either a DW for $w=1/(d+2)$ or a CL brane for $w\neq1/(d+2)$. The relation~\eqref{rel}
implies that the free energy of the dual QFT is constant. This is what we expect because, owing to the absence of solutions
with event horizons, our brane solution does not allow BB excitations at finite temperature.

The null energy conditions for the bulk stress-energy tensor require~\cite{Dong:2012se}
\be\lb{ppq}
(d-\theta)\left(d (z-1)-\theta\right)\ge0,\quad (z-1) (d+z-\theta)\ge0.
\ee%
Taking into account that the constraint $0\le{}w\le1$ implies the condition $0\le{}z\le{}2d/(d-1)$, it is straightforward to check that
the first inequality is always satisfied. The second one is saturated in our case, as expected because the source is a massless field.

\subsection{Electrically charged brane solutions in four dimensions\lb{sect:s3_4}}

Let us now consider the case of non-vanishing electric charge. Taking for simplicity $d=2$, \cref{z3LL} become
\be\label{zz11}
Y'+Y^2=-{(\phi')}^2,\quad
(u\phi')'=0,\quad
u''- 4(uY)'=4Q^2 e^{-2\int{}Y},\quad
u''=0.
\ee%
Again, we first determine $u$ and $\phi$, then we solve the Riccati equation, and finally we 
use the third equation to express the charge $Q$ in terms of the integration constants:
\be\lb{solcha1}
U=\left(\frac{r}{r_0}\right)^{1 -2w}\left[1-c_1\left(\frac{r}{r_0}\right)^{1 -2w}\right]^{-2},\quad
R^2=\left(\frac{r}{r_0}\right)^{2w}\left[1-c_1\left(\frac{r}{r_0}\right)^{1 -2w}\right]^2,\quad
\phi=-\gamma\log\left(\frac{r}{r_0}\right)+ \phi_0,
\ee%
where $w-w^2=\gamma^2$ and $c_1=r_0^2Q^2/(1 -2w)^2$. These constraints imply $0\le{}w\le1$, with $w\neq 1/2$.

Solutions~\eqref{solcha1} represent a three-parameter ($r_0,\,w,\,Q$) family of brane solutions, with a naked singularity at
$r=r_0c_1^{(1/2w-1)}$ for all $w$ in the range $0\le{}w\le1$. In fact the scalar curvature is 
\be\lb{sc1}
\mathcal{R}=\frac{2\gamma^2}{r_0^2}\left(\frac{r}{r_0}\right)^{-1 -2w}\left[1-c_1\left(\frac{r}{r_0}\right)^{1 -2w}\right]^{-2}.
\ee%

For $w=0,1$ the solutions reduce to curved branes with vanishing scalar curvature $\mathcal{R}=0$, sourced by a constant
scalar field. In particular one has:
\be%
U=\frac{r}{r_0}\left(1-\frac{c_1 r}{r_0}\right)^{-2},\quad R=1-\frac{c_1 r}{r_0} \quad\text{for } w=0,\quad
U=\frac{r_0}{r}\left(1-\frac{c_1 r_0}{r}\right)^{-2},\quad R=\frac{r}{r_0}-c_1 \quad\text{for } w=1.
\ee%
Both  solutions represent a sort of Reissner-Nordstr\"om (RN) branes. In fact, the metric part of  the solutions can be written
using a translation and a rescaling of the radial coordinate $r$ in the form $ U= \frac{\alpha}{r}+ \frac{Q^2}{r^2},\, R=r$,
where $\alpha$ is a constant.

The asymptotic behavior of solution~\eqref{solcha1} in general is ruled by the value of $w$. When $0<w<1/2$,
we have for $r\rightarrow\infty$ (corresponding to $\phi\rightarrow-\infty$) at leading order: 
\be\label{asy1}
U= c_1^{-2}\left(\frac{r}{r_0}\right)^{-1+2w},\quad R^2=c_1^2\left(\frac{r}{r_0}\right)^{2-2w},\quad\phi=-\gamma\log\frac{r}{r_0},
\ee%
while when $1/2<w<1$ one finds:
\be%
U=\left(\frac{r}{r_0}\right)^{1-2w},\quad R^2=\left(\frac{r}{r_0}\right)^{2w}\label{asy2},\quad\phi=-\gamma\log\frac{r}{r_0}.
\ee%

Notice that in this last case the asymptotic behavior coincides with the uncharged solution~\eqref{sol31} with $d=2$,
discussed in the previous subsection. It is also easy to check that the two asymptotic forms~\eqref{asy1} and~\eqref{asy2}
are mapped one into the other by the transformation $w\rightarrow1-w$, that leaves invariant the constraint $w-w^2=\gamma^2$.

\subsubsection{Energy of the solution}

Also in this case, we can compute the total energy of the solution using the Euclidean action formalism~\cite{Martinez:2006an}.
The variation of the boundary terms of the action are:
\be\label{mass1}
\delta{}M =\delta{}M_G+\delta{}M_\phi-16\pi\,Q\,\delta\Phi|^\infty,
\ee%
where $\delta{}M_G$ and $\delta{}M_\phi$ are the gravitational and scalar contributions, corresponding to Eq.~\eqref{mass}
with $d=2$, while the last term is the contribution due to the charge and $\Phi$ is the electric potential.

Evaluating the energy for our solution~\eqref{solcha1}, we find that the electromagnetic contribution always vanishes at $r=\infty$,
while the full result is ruled by the value of $w$ again. When $0\le w\le 1/2$ we obtain
\be\label{mass11}
M=\frac{16\pi}{r_0}[- w + (2w-1)\log(Q/Q_0)],
\ee%
where $Q_0$ is an integration constant. In this case the sign of the energy depends on the mutual values of $w$ and $Q$.  
When $1/2\le w\le 1$ (where the asymptotic behavior of the charged brane~\eqref{solcha1} coincides with the uncharged brane~\eqref{sol31} for $d=2$),
one simply finds 
\be\label{mass12}
M=-\frac{16\pi}{r_0}.
\ee%
As expected, it coincides with the energy~\eqref{mass0} of the uncharged solution with $d=2$.

\subsubsection{Scaling symmetries and hyperscaling violation}

Let us now study the scaling symmetries of the solution~\eqref{solcha1} in the UV regime, \ie\ for $r\to\infty$. 

We consider first the asymptotic form~\eqref{asy1}, describing the UV regime of the solution when $0\le w\le 1/2$.
Let $\tilde{r}_0= 2r_0c_1$, then~\eqref{asy1} can be put in the form~\eqref{hv} by the transformation of coordinates
\be%
\frac{r}{r_0}\to\left(\frac{r}{\tilde{r}_0}\right)^2,
\ee%

After this transformation, it is  simple to extract the hyperscaling violation parameter and the dynamical exponent:
\be\label{cp}
z = 4-4w, \quad \theta=6-4w.
\ee%

From \cref{cp} one can  sees that  $\theta=0$ for $w=3/2$, which is outside the range  of $w$. Hence, our brane solution~\eqref{solcha1}
cannot describe neither an AdS nor a Lifshitz brane, \ie\ it never belongs to the subclasses (\ref{item1}) or (\ref{item2}) of \cref{sect:hyper}.
The brane is either a DW for $w=3/4$ or a CL brane for $w\neq3/4$. Notice also that the relation $\theta=z+2$
(already observed in the uncharged case in its $d$-dimensional form), is verified again, as well as the constant free energy of the dual QFT\@.

The null energy conditions for the bulk stress-energy tensor
\be%
(2-\theta)[2 (z-1)-\theta]\ge0,\quad (z-1) (2+z-\theta)\ge0
\ee%
are satisfied. In particular, the second one is saturated.

Exploiting the symmetry of the  asymptotic solutions~\eqref{asy1} and~\eqref{asy2} under $w\to1-w$ described in \cref{sect:s3_4},
we can easily derive the critical exponents $\theta$ and $z$ related to the scaling~\eqref{asy2}, which, as expected, 
are those of~\eqref{para} with $d=2$.

\section{Classifications of brane solutions  sourced by scalar fields\lb{sect:s4}}

In this section we  present a detailed  physical  classification of brane solutions sourced by a scalar field with 
a self-interaction potential $V$, \ie\ a classification of the brane solutions of the  MCEMSG theory~\eqref{action}.
Obviously, the form of the brane solution will strongly depend on the form of the potential $V$ and a complete classification is 
without reach. On the other hand, having in mind holographic applications, we are not interested in generic solutions  but  on  branes 
that respect some scaling symmetries, at least  asymptotically ($r\to\infty$) and in the $r=0$ region. We can therefore  built up a brane 
classification based on the scaling symmetries  discussed in \cref{sect:hyper}.

We will preliminary show   that  the scale-covariant solution~\eqref{sol31} together with the  solutions already known in the
literature give all the possible brane solutions of MCEMSG of the form~\eqref{hv}.
Using  a reparametrisation of the radial coordinate $r$ we can easily write  the metric~\eqref{hv} in the form~\eqref{pmetric}
with $U=B(r/r_0)^\alpha,\, R=D(r/r_0)^\beta$, from which it follows $Y=\beta/r$. Inserting this expression of $Y$ in the field
equations~\eqref{zzll} with $Q=0$ one finds only three classes of solutions:
\begin{enumerate}[(I)]
\item $\beta=1,\,\phi=\const,\,u\propto{}r^{d+2},\,V=-\Lambda^2$, which corresponds to the $\AdS_{d+2}$ brane;
\item $\phi\propto\log(r/r_0),\,u\propto{}r,\,V=0$, which corresponds to the solution~\eqref{sol31};
\item $\phi\propto\log(r/r_0),\,u\propto{}r^\eta,\,V\propto e^{\mu\phi},\,\eta\neq\{1,d+2\}$, which corresponds to
DW solutions sourced by an exponential potential~\cite{Cadoni:2011nq}.
\end{enumerate}
In our classification, we will distinguish between \emph{elementary} solutions, \ie\ solutions respecting some  scaling symmetry,
and \emph{interpolating} solutions, \ie\ solutions that approach to elementary branes only in the $r=0$ IR  region and in the $r=\infty$ UV region.
We will discuss separately  these two types of solutions.

\subsection{Elementary solutions\lb{sect:es}}

Elementary solutions are defined as those solutions of the theory~\eqref{action} which belong to one of the subclasses
of \cref{sect:hyper}. In principle, we should therefore have  in correspondence with the  four scale
symmetries of \cref{sect:hyper} four kinds of elementary branes. However, as a consequence of the previous demonstration, the
Lifshitz solution~(\ref{item2}) cannot be obtained if the source is a minimally coupled scalar field. We are therefore left with three classes of solutions:
\begin{enumerate}[(A)]
\item\lb{caseA} \emph{AdS branes} are $Q=0$ solutions of the model~\eqref{action} when the potential $V(\phi)$ is a negative cosmological
constant or has a local extremum $V'(\phi_0)=0$, with $V(\phi_0)= -\Lambda^2$. In this case we have a trivial (constant) scalar field~$\phi=\phi_0$.
\item\lb{caseB} \emph{Domain Walls} are $Q=0$ solutions of the model~\eqref{action} when the potential $V(\phi)$ is a pure exponential:
$V(\phi)\propto e^{\mu\phi}$. DWs are sourced by a scalar behaving logarithmically: $\phi\propto \log r/r_0$~\cite{Cadoni:2011nq}.
\item\lb{caseC} \emph{Conformal-Lifshitz branes}  are solutions of the model~\eqref{action}  in the case $V=0$  for $Q=0$. For $Q\neq 0$ they
appear as solution of the theory for a purely exponential potential~\cite{Cadoni:2011nq}.
\end{enumerate}
For $z$ and $\theta$ in \cref{hv} finite, there are no other elementary brane solutions which can be sourced by a minimally coupled scalar field.
However, in the $\theta=0,\, z\to\infty$  limit, it is known that the Lifshitz brane becomes the  $\AdS_2\times R_d$ spacetime, see \eg~\refcite{Bhattacharya:2012zu}.
If one wants to include this limiting case one should consider also  a fourth kind of elementary brane solutions, namely   \emph{$\AdS_2\times R_d$ branes}.
These spacetimes  are $Q\neq 0$ charged  solutions of the model~\eqref{action}  when the potential $V(\phi)$ is a negative cosmological constant
or has a local extremum $V'(\phi_0)=0$, with $V(\phi_0)= -\Lambda^2$. Similarly to case~(\ref{caseA}) these branes are sourced by  a constant  scalar field.

\subsection{Interpolating solutions\lb{sect:is}}

Combining the three  types of elementary brane solutions discussed above one can construct different kinds
of interpolating solutions, \ie\ solutions  behaving only in the UV and  IR regimes as an elementary brane. 
The interpolating solutions are very useful for holographic applications,  in particular for AdS/CFT and the gravity/condensed
matter correspondence of EMSG~\cite{Cadoni:2009xm,Charmousis:2009xr,Goldstein:2009cv,Gubser:2009qt,Cadoni:2011yj,Cadoni:2011nq,%
Cadoni:2012ea,Gouteraux:2011qh,Bertoldi:2011zr,Cadoni:2012ea,Gouteraux:2012yr,Cadoni:2012uf,Cadoni:2013hna}.

The recent literature dealing with  this topic contains a multitude of such interpolating  brane  solutions derived
in the context of the gravity  theory~\eqref{action} and its possible generalizations (covariant coupling between  the $\textrm{U}(1)$
gauge field $A_\mu$ and the scalar, coupling between the Maxwell tensor $F_{\mu\nu}$ and the scalar,  Einstein-Yang-Mills-scalar gravity, etc.).
Despite this variety of solutions and models, the simplest case described by the action~\eqref{action} is extremely important
for the crucial role played by the scalar field. In the dual QFT the scalar field  gives an order parameter triggering symmetry
breaking and/or phase transitions. Moreover, $\phi$ has a nice interpretation in terms of holographic renormalization group equations
describing the flow between UV/IR fixed points, see \eg~\refcite{deBoer:2000cz}.

The classification of the possible interpolating solutions of the theory~\eqref{action} is simple because it is parametrized by
a single function, the potential $V(\phi)$. It follows that the interpolating solutions are essentially determined by the behaviour
of the potential in the IR and UV region. This feature is not present in other, more complicated, models in which the presence of two
or more coupling functions prevents a simple classification.  In the following we will list all the known interpolating solutions and,
in the case they are not been already discussed in the literature, we will discuss their possible existence.
\begin{description}
\item[AdS-AdS interpolating solutions]
In general, solutions of this kind are present when the potential has a local maximum and a local minimum
connected with  continuity. The gravitational soliton bridges  two AdS spacetimes whereas the
dual field theory flows from a fixed point in the UV to an other fixed point in the IR\@.
The two CFTs are connected by the c-theorem, which  gives  well-defined predictions for 
the running of the central charge when running from the UV to the IR\@.
Interpolating solutions of this kind are typically  numerical solutions and   have been 
already discussed  in the literature, see \eg~\refcite{Gubser:2009gp}.
\item[AdS-DW  Interpolating solutions]
Typically, these  solutions are present when $Q=0$ and the potential  has an extremum at $\phi_0$ with $V(\phi_0)<0$  in  the UV (IR), 
whereas it behaves exponentially in the IR (UV). The gravitational soliton interpolates between an AdS spacetime at $r=\infty$ ($r=0$)
and  a DW near $r=0$ ($r=\infty$). The dual QFT flows from a fixed point in the UV (IR) to an hyperscaling violating phase in
the IR (UV). Exact solutions of this kind are known, both in the case of hyperscaling violation in the IR~\cite{Cadoni:2011nq} and
hyperscaling violation in the UV~\cite{Cadoni:2011yj,Cadoni:2012uf}. Several numerical solutions are also known, see \eg~\refcite{Cadoni:2013hna}.
\item[AdS-CL  Interpolating solutions]
Brane solutions of the theory~\eqref{action} bridging an AdS spacetime in the UV (IR)  with a CL solution in the IR (UV) have not been
discussed in the literature. Conversely, they are quite common in non-minimally coupled theories and in the case of  holographic
superconductors.  In the context of the minimally coupled theory they are expected to show up  in two cases: first, $V(\phi)$ has an
extremum in the UV (IR)  whereas in the IR (UV) region the kinetic energy  of the scalar dominates over its potential energy so that
we can use $V\sim 0$; second, we have $Q\neq 0$ charged solutions, $V(\phi)$ has  an extremum in the UV (IR)  whereas in the IR (UV)
region $V$ behaves exponentially. Obviously the existence of these solutions has to be checked numerically.
\item[DW-DW Interpolating solutions]
Solutions interpolating between two DW branes are not known in literature. However, we can easily find  a form of the potential which
is a good candidate  for  generating  this kind of solution. One  can start from a simple combination of exponentials:
$V (\phi)= A e^{\alpha \phi} + B e^{-\beta \phi}$, that obviously  behaves as a single exponential in the two regimes $\phi\to\infty$ and
$\phi\to-\infty$. We know that a simple exponential form of the potential leads, in the case of uncharged branes, to a DW solution~\cite{Cadoni:2011nq}.
Thus, the corresponding  brane solutions of the model, if existing would give a  soliton interpolating between two DWs at $\phi=\pm\infty$.
\item[DW-CL  Interpolating solutions]
Having in mind the features of the elementary solutions discussed in \cref{sect:es}, one can expect this kind of solution to show up in
the case of a potential which diverges exponentially in a region whereas  approaches to zero in an other region. Solutions of this type,
although already known in the literature~\cite{Cadoni:2011nq}, have not been recognized as DW-CL interpolating solutions. We will show in \cref{sect:s4_1}
that for an appropriate choice of the parameters the solutions of \refcite{Cadoni:2011nq} describe a DW-CL interpolating solution.
\item[CL-CL  Interpolating solutions]
In order to generate these kind of branes one should consider a potential which vanishes in two distinct regions. Alternatively one can
consider charged solutions and a potential behaving  exponentially. Also a mixed charged configuration, with a  potential vanishing in one
region and behaving exponentially in an other region, is possible. Also solutions of this type are known in literature~\cite{Cadoni:2011nq},
but have not been recognized as CL-CL interpolating solutions. We will show in \cref{sect:s4_2} that for an appropriate choice of the parameters the solutions
of \refcite{Cadoni:2011nq} describe a CL-CL interpolating solution.
\end{description}

In the  above classification of interpolating brane solutions we have not considered the limiting case in which one of the elementary
solution is $\AdS_2\times R_d$. Brane solutions interpolating between an elementary solution (\ref{caseA}), (\ref{caseB}) or (\ref{caseC}) in the 
UV and  $\AdS_2\times R_d$ in the IR possibly exist whenever one considers charged branes and a potential $V$ behaving in the IR as 
a negative cosmological constant.  The simplest, well-known, example of this kind is obtained considering $V=-\Lambda^2$ identically.
The charged brane solutions are simply given by the AdS-RN BB\@. In the extremal limit, when the BPS bound is saturated,
we get a solitonic solution which interpolates between $\AdS_{d+2}$ in the $r\to\infty$ region and $\AdS_2\times R_d$ in the near-horizon region.
The $\AdS_2\times R_d$ geometry and related interpolating solutions are of interest also because they may act as IR regulators of
the generic scale-covariant geometry~\eqref{hv}~\cite{Bhattacharya:2012zu}.

All the above interpolating solutions are considered as branes without event horizons, \ie\ as zero temperature solutions.
An important question, particularly in view of holographic applications, is if they can be considered as the extremal limit of BB
solutions with non-trivial hair, \ie\ solutions at finite temperature endowed with a non-trivial scalar field.
There is no general answer to this question. Owing to no-hair theorems~\cite{Israel:1967wq,Bekenstein:1995un,Hertog:2006rr}
the existence of hairy solutions is related to global features of the potential $V(\phi)$.
Nevertheless, in  most examples discussed in the literature the interpolating solutions appear as extremal limit of BB solutions.

\section{Domain-Wall/Conformal-Lifshitz interpolating solutions\lb{sect:s4_1}}

In this section we  discuss exact solutions, which interpolate between a DW and a CL brane. In the previous section we have
seen that this kind of solution requires uncharged branes and  a potential which diverges exponentially in a region, whereas
approaches to zero in an other region. We are therefore lead to consider $Q=0$ solutions and  the following simple potential ($0\le w\le 1$)
\be\lb{kk2}
V (\phi)= A d w \left(1- w (d+2)\right) e^{-\sqrt{\frac{8(1-w)}{dw}}\,\phi}, 
\ee%
which diverges exponentially for $\phi\to -\infty$, while $V\to 0$ for $\phi\to \infty$.
The  general solution for the theory~\eqref{action} with this potential  is given by~\cite{Cadoni:2011nq}
\be\lb{solcha2}
U= A R^2-c \left(\frac{r}{r_0}\right)^{1-dw},\quad
R^2=\left(\frac{r}{r_0}\right)^{2w},\quad
\phi= \sqrt{\frac{d}{2} (w-w^2)}\log\left(\frac{r}{r_0}\right),\quad 0\le{}w\le1
\ee%
where $c$ is an integration constant. Notice that when $A=0$ and $c<0$ the solution~\eqref{solcha2} becomes exactly 
the solution~\eqref{sol31}. This is consistent with the fact that for $A=0$, the potential is identically zero and the
solution is sourced  by the kinetic term of the scalar.

When $c>0$ and for $1/(d+2)<w\le 1$, the solution~\eqref{solcha2} describes a BB with  an event horizon~\cite{Cadoni:2011nq}.
On the other hand, when $c<0$  the solution has no horizon and   depending on the value of $w$ it has different asymptotic
behaviour. For $0\le w<1/(d+2)$  at $r\to\infty$   the second term in the metric function $U$  dominates over the first.
Asymptotically at $r\to\infty$ (corresponding to $\phi\to\infty$) $V$ approaches to zero and the solution becomes the
CL brane solution discussed in the previous section. Conversely, near $r=0$ (corresponding to $\phi=-\infty$)  $V$ 
diverges,  the first term dominates over the second and the solution becomes a DW\@. Thus the global solution~\eqref{solcha2}
interpolates between a  CL brane in the UV and a DW in the IR\@. Physically, this means that the UV behaviour is dominated
by the kinetic energy of the scalar field, whereas the IR behaviour is dominated by the potential energy of the scalar.
Obviously, for $c<0$ and $1/(d+2)<w\le 1$  the picture is reversed and we have a global solution interpolating between a CL in the IR and a DW in the UV\@.

For $c>0$, solution~\eqref{solcha2} describes a BB and we can associate to it thermodynamical parameters.
Using \cref{mass} we can compute the total mass of the BB\@. In particular for the range of parameters for which
the solution describes a BB (namely $c>0$ and $1/(d+2)<w\le 1$), we find: $M=4\pi Adwc/r_0$.
The temperature $T$ and the entropy $S$ of the BB solution can be calculated using the well-known formul\ae\
$T=U'(r_{h})/4\pi$ and $S=16\pi^2 R^d(r_{h})$:
\be\label{tempentropy}
T=\frac{A[(d+2)w-1]}{4\pi{}r_0}c^\frac{2w-1}{(d+2)w-1},\quad S=16\pi^2 c^\frac{dw}{(d+2)w-1}.
\ee%
Using these equations it is easy to verify that the first principle $dM=T\,dS$ is satisfied.
Notice that in the extremal limit $c=0$ the BB becomes, as expected in view of its IR behaviour, a DW\@.

\section{Conformal-Lifshitz/Conformal-Lifshitz  interpolating solutions\lb{sect:s4_2}}

In \cref{sect:is} we have seen that CL-CL interpolating solutions require charged branes ($Q\neq 0$) and a potential having the same qualitative behaviour of~\eqref{kk2}.
Considering for simplicity  the four-dimensional case, $d=2$,  we take  a non-vanishing electric charge and the potential
\be\lb{kk3}
V (\phi)= \frac{2Q^2(1-w)}{1-3w}e^{-4\sqrt{\frac{w}{1-w}}\,\phi}.
\ee%

The general brane  solutions are given by~\cite{Cadoni:2011nq}:
\be\label{solcha3}
U &= B \left(\frac{r}{r_0}\right)^{2-4w}\left[1-C \left(\frac{r}{r_0}\right)^{-1+2w}\right],\quad 
R^2=\left(\frac{r}{r_0}\right)^{2w},\quad
\phi= \sqrt{w-w^2}\log\left(\frac{r}{r_0}\right),\\
B &=\frac{2Q^2 r_0^2}{(1-2w)(1-3w)},\quad 0\le{}w\le1\0.
\ee%

Also in this case, for $C>0$ and $0\le w<1/2$ (with $w\neq1/3$) the solution~\eqref{solcha3} describes a BB with an event horizon.
For $C<0$   the brane has  no horizon and   we have a brane interpolating between two CL elementary solutions in the IR and UV regions.
When $C<0$ and $w>1/2$, for  $r\to\infty$ ($\phi\to\infty$) the potential approaches to zero and the solution  gives an elementary
CL solution, which coincides with the $r\to\infty$ (and $w>1/2$) regime of the electrically charged  brane~\eqref{asy1}.
Also near $r=0$ the solution reduces to an elementary CL  brane, but with a different dynamical exponent. 
For $C<0$ and $0\le w<1/2$ ($w \neq 1/3$) we have the same limiting elementary CL branes but with the IR and UV regions exchanged.

When the solution~\eqref{solcha3}  describes BBs ($C>0$) we can compute the associated  thermodynamical parameters:
\be%
M=\frac{8\pi B w}{r_0}\,C,\quad T=\frac{B(1-2w)}{4\pi r_0}C^\frac{1-4w}{1-2w},\quad S=16\pi^2 C^\frac{2w}{1-2w},
\ee%
and check that the first principle $dM=T\,dS$ is satisfied.
It is interesting to notice that  the $C=0$ extremal limit  of these BB solutions is described by a CL brane.

\section{Conclusions\lb{sect:concl}}

In this paper we have derived brane solutions of MCEMSG in $d+2$ dimensions in the case
of a vanishing potential. We have shown that these brane solutions belong to the broad class of scale-covariant
metrics, which generate hyperscaling violation in the holographically dual QFT\@. Moreover, these solutions can
be considered as the brane counterpart of the well-known JNWW, spherically symmetric, solutions of Einstein-scalar gravity with $V=0$.
We have also explicitly shown that our brane solution, together with the AdS brane and the DW solution sourced by an 
exponential potential, give all the possible scale-covariant, hyperscaling violating, geometries of MCEMSG\@.
Using this result we have been able to give a comprehensive and detailed classification of the brane solutions of
the theory in terms of  symmetric (elementary) and interpolating solutions, which can be very useful for holographic
applications. In particular, the interpolating solutions can find a broad field of holographic applications because
the dual QFT describes the flow from different regimes (fixed points, hyperscaling violation, Lifshitz) in the UV and IR,
characterized  by different scaling symmetries. In this context is important to stress the fact that some of our solutions
have curvature singularities at $r=0$ (in the IR of the dual QFT). In this cases an IR completion of the theory is needed. 
From the bulk point of view these completion  can be realized using an IR regular geometry such as $\AdS_{d+2}$~\cite{Cadoni:2012ea}
or $\AdS_2\times R_d$~\cite{Bhattacharya:2012zu}.

\appendix
\section{Spherically symmetric solutions in four dimensions\lb{app:chargedJNWW}}

This paper has been focused on brane solutions, \ie\ on solutions  with $\varepsilon=0$  in the field equations~\eqref{fed}.
However, our method for solving the field equations described in \cref{sect:s3} can be also used to derive solutions having
$d$-dimensional sections with spherical topology, \ie\ solutions with $\varepsilon=1$, sourced by  a scalar field with vanishing potential.
For spherically symmetric solutions in $d=2$, \cref{z3LL} become
\be\label{zz1}
Y'+Y^2=-{(\phi')}^2,\quad
(u\phi')'=0,\quad
u''- 4(uY)'=-2+4Q^2 e^{-2\int{}Y},\quad
u''=2.
\ee%

The uncharged $Q=0$ solution of the previous equation gives the well-known JNWW solution~\cite{Janis:1968zz,Wyman:1981bd}.
The derivation of this solution using field equations in the form~\eqref{zz1} has been already discussed in \refcite{Cadoni:2015gfa}. 
For the $Q\neq 0$, charged case we solve \cref{zz1} by determining first $u$ and $\phi$ then by solving the Riccati equation.
Finally, we use the third equation to express the $Q$ in terms of the integration constants. We have
\bsube[solcha]
U&={\left(1+\frac{r_0}{r}\right)}^{2w-1}{\left[1-c_1\left(1+\frac{r_0}{r}\right)^{2w-1}\right]}^{-2},\quad
R^2=r^2{\left(1+\frac{r_0}{r}\right)}^{2(1-w)}{\left[1-c_1{\left(1+\frac{r_0}{r}\right)}^{2w-1}\right]}^2,\\
\phi&=-\gamma\log\left(1+\frac{r_0}{r}\right)+\phi_0,\quad
w-w^2=\gamma^2,\quad c_1=\frac{Q^2}{r_0^2 (1 -2w)^2}.
\esube%

For $w\neq0,1$, these solutions can be considered as the charged generalization of the 
JNWW solutions and describe a spacetime with a naked singularity at $r=r_0\left(c_1^{1/(1-2w)}-1\right)^{-1}$.
In fact the scalar curvature of the spacetime is given by:
\be\lb{scca}
\mathcal{R}=\frac{2\gamma^2 r_0^2}{r^4}\left(1+\frac{r_0}{r}\right)^{2w-3}\left[1-c_1\left(1+\frac{r_0}{r}\right)^{2w-1}\right]^{-2}.
\ee%
For $w=0,1$ the solution gives the usual RN black hole solution with a constant scalar field $\phi$.
In fact in these cases the previous solution can be put in the usual RN form by rescaling and translating the radial coordinate~$r$.

\begin{acknowledgments}
EF acknowledges financial support provided under the European Union's H2020 ERC Consolidator Grant
``Matter and strong-field gravity: New frontiers in Einstein's theory'' grant agreement no.~MaGRaTh-646597.
\end{acknowledgments}

\bibliographystyle{apsrev41b}
\bibliography{references}

\end{document}